\documentclass[sigconf]{acmart}

\AtBeginDocument{%
  }

\setcopyright{acmlicensed}
\copyrightyear{2026}
\acmYear{2026}
\setcopyright{cc}
\setcctype{by-nc-nd}
\acmConference[WWW '26]{the ACM Web Conference 2026}{April 13-17, 2026}{Dubai, United Arab Emirates}
\acmBooktitle{Proceedings of the ACM Web Conference 2026 (WWW '26), April 13--17, 2026, Dubai, United Arab Emirates}
\acmISBN{979-8-4007-2307-0/2026/04}
\acmDOI{10.1145/3774904.3792446}

\usepackage{amsmath,amsfonts,bm}

\def\eqref#1{equation~\ref{#1}}

\def\1{\bm{1}}

\DeclareMathAlphabet{\mathsfit}{\encodingdefault}{\sfdefault}{m}{sl}
\SetMathAlphabet{\mathsfit}{bold}{\encodingdefault}{\sfdefault}{bx}{n}

\def\sR{{\mathbb{R}}}

\newcommand{\E}{\mathbb{E}}

\DeclareMathOperator*{\argmax}{arg\,max}

\newcommand*{\dif}{\mathop{}\!\mathrm{d}}

\usepackage{thm-restate, amsthm, algorithm, algorithmicx, varwidth}
\usepackage[noend]{algpseudocode}

\usepackage{balance}

\algnewcommand{\LineComment}[1]{\State \(\triangleright\) #1}

\newtheorem{defn}{Definition}
\newtheorem{cor}{Corollary}
\newtheorem{rem}{Remark}
\floatname{algorithm}{Mechanism}

\begin{document}

\title{Pay for The Second-Best Service: A Game-Theoretic Approach Against Dishonest LLM Providers}

\author{Yuhan Cao}
\authornote{Equal contribution.}
\email{yuhanteafrog@gmail.com}
\affiliation{%
  \institution{Shanghai Qi Zhi Institute}
  \city{Shanghai}
  \country{China}}
\affiliation{%
  \institution{ShanghaiTech University}
  \city{Shanghai}
  \country{China}}

\author{Yu Wang}
\email{yuuwwang@gmail.com}
\authornotemark[1]
\affiliation{%
  \institution{Institute of Information Engineering, Chinese Academy of Sciences}
  \city{Beijing}
  \country{China}}
\affiliation{%
  \institution{Shanghai Qi Zhi Institute}
  \city{Shanghai}
  \country{China}}

\author{Sitong Liu}
\email{liust2023@shanghaitech.edu.cn}
\affiliation{%
  \institution{ShanghaiTech University}
  \city{Shanghai}
  \country{China}}

\author{Miao Li}
\email{limiao@bigai.ai}
\affiliation{%
  \institution{Beijing Institute for General Artificial Intelligence, BIGAI}
  \city{Beijing}
  \country{China}}

\author{Yixin Tao}
\email{taoyixin@mail.shufe.edu.cn}
\affiliation{%
  \institution{Shanghai University of Finance and Economics}
  \city{Shanghai}
  \country{China}}

\author{Tianxing He}
\email{hetianxing@mail.tsinghua.edu.cn}
\authornote{Corresponding author.}
\affiliation{%
  \institution{Shanghai Qi Zhi Institute}
  \city{Shanghai}
  \country{China}}
\affiliation{%
  \institution{Tsinghua University}
  \city{Beijing}
  \country{China}}
\affiliation{%
  \institution{Xiongan AI Institute}
  \city{Hebei}
  \country{China}}

\renewcommand{\shortauthors}{Yuhan Cao et al.}

\begin{abstract}
  The widespread adoption of Large Language Models (LLMs) through Application Programming Interfaces (APIs) induces a critical vulnerability: the potential for dishonest manipulation by service providers. This manipulation can manifest in various forms, such as secretly substituting a proclaimed high-performance model with a low-cost alternative, or inflating responses with meaningless tokens to increase billing. This work tackles the issue through the lens of algorithmic game theory and mechanism design. We are the first to propose a formal economic model for a realistic user-provider ecosystem, where a user can iteratively delegate $T$ queries to multiple model providers, and providers can engage in a range of strategic behaviors. As our central contribution, we prove that for a continuous strategy space and any $\epsilon\in(0,\frac12)$, there exists an approximate incentive-compatible mechanism with an additive approximation ratio of $O(T^{1-\epsilon}\log T)$, and a guaranteed quasi-linear second-best user utility. We also prove an impossibility result, stating that no mechanism can guarantee an expected user utility that is asymptotically better than our mechanism. Furthermore, we demonstrate the effectiveness of our mechanism in simulation experiments with real-world API settings.
\end{abstract}



\maketitle

\section{Introduction}

\begin{figure*}[!ht]
    \centering
    \includegraphics[width=\linewidth]{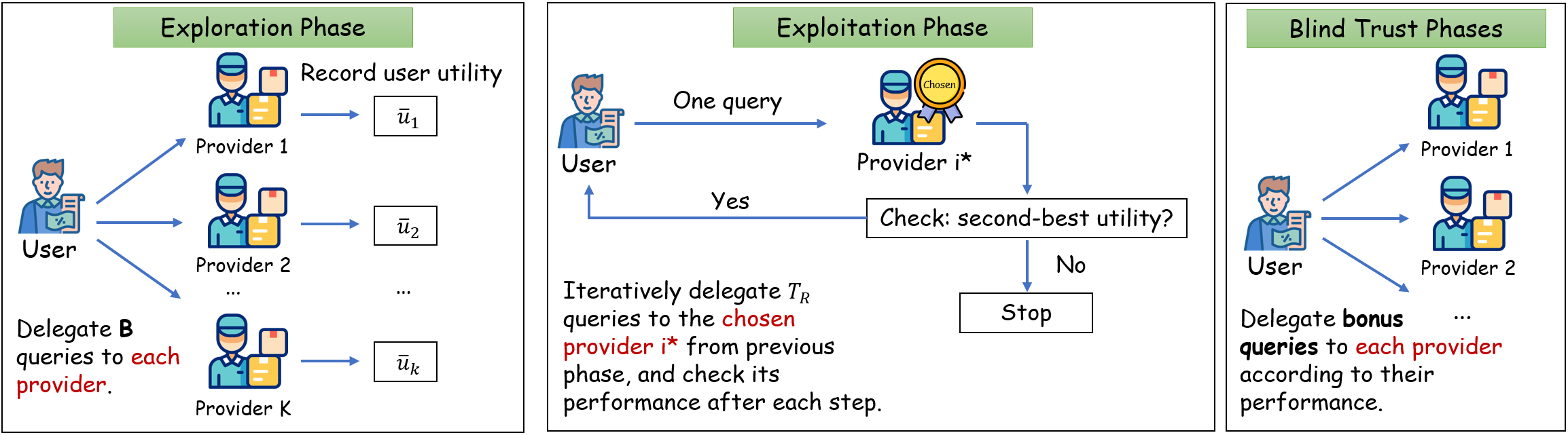}
    \caption{An illustration figure of our proposed mechanism, in which the provider's optimal strategy would guarantee the user a second-best overall utility. The mechanism splits the user's queries among four sequential phases. In the first phase (exploration), a batch of sample outputs from each service provider is collected. They need to compete to get selected in the second phase (exploitation), where the best-performing provider is asked to deliver the second-best user utility for the majority of queries. This is followed by the final two blind trust phases (detailed in Section~\ref{sec:mech}), which serve as incentives for providers to behave well in the first two phases.}
    \label{fig:mechanism}
\end{figure*}

In recent years, large language models (LLMs) have demonstrated capabilities across a multitude of domains \citep{thirunavukarasu2023large,hong2024metagpt, balunovic2025matharena}. Following the release of ChatGPT \citep{chatgpt2022}, there has been a rapid and widespread adoption of LLMs among the general public, embedding them into a variety of commonplace activities. The inference processes of these models are predominantly controlled by companies, which provide access to individual users through interactive interfaces and Application Programming Interfaces (APIs). The growing user base has also triggered the development of a secondary market (e.g., TogetherAI \citep{togetherai2023} and OpenRouter~\citep{openrouter2025}), where providers aggregate resources to offer more affordable services.

This market expansion introduces a critical vulnerability: the potential for service provider dishonesty. Since users interact with LLMs in a ``black-box'' manner, a clear economic incentive exists for providers to substitute the advertised models with less capable alternatives to reduce their own operational costs. Such deceptive practices can manifest in various forms. For instance, when user is asking for a large model like Qwen3-235B-A22B \citep{yang2025qwen3technicalreport}, a provider might secretly switch to a smaller one, such as Qwen3-30B-A3B. It might also deploy a quantized version (e.g., using INT8 or FP8) \citep{gao2025model}. Additionally, providers might dynamically adjust a model's reasoning effort (as in reasoning models like OpenAI o1~\citep{o12024}) without user consent. The pay-per-token API billing also introduces new risks, allowing providers to inflate charges by inserting superfluous tokens that are difficult for users to detect \citep{kumar2025overthinkslowdownattacksreasoning}.

These dishonest behaviors have profound negative implications. Firstly, they erode the trustworthiness of the service and promote vicious competition within the market. Secondly, for researchers dependent on APIs for scientific investigation, undisclosed behaviors severely undermine experimental reproducibility.

Existing research has approached this challenge from a technical detection standpoint. Efforts include auditing model outputs for semantic behavioral drift \citep{chen2023chatgptsbehaviorchangingtime, 10.1145/3630106.3659047} or token-level consistency \citep{miao-fang-2025-user, sun2025coincountinginvisiblereasoning}, and examining the inherent discrepancy across different models \citep {gao2025model, sun2025idiosyncrasieslargelanguagemodels, cai2025gettingpayforauditing}. In this work, we tackle the problem through the lens of algorithmic game theory and computational economics. In other words, we propose a mechanism in which the providers would compete with each other, thus protecting user utility.

\subsection{Our Contributions}

\paragraph{The User-Provider Delegation Game Model.} We model an online user-provider delegation game within a market consisting of a single user and multiple candidate providers, each declaring a specific LLM for service. The user has $T$ sequential queries. For each query, the user selects and delegates it to one provider. Upon accepting the delegation,  a strategic provider could take action by choosing a model with a different actual cost or reporting an inflated output token sequence. After receiving the token sequence and making a payment, the user evaluates the utility gained, which informs their future delegation decisions.

We formulate a mechanism design problem in which providers determine their strategies based on the user's delegation mechanism. We characterize the providers' action space, which comprises cost and token sequence reporting. Our proposed model, which draws upon both mechanism design theory \citep{nisan1999algorithmic} and contract theory \citep{10.1561/0400000113}, addresses a novel problem that has not been extensively studied in prior literature.
To better study the relationship between provider actions and outcomes, we also model the performance of LLMs. This specifies how cost control affects the expected reward and output sequence length.

\paragraph{Impossibility Result and Utility Guarantee Mechanism.}
We provide a theoretical guarantee under the continuous cost control model. Inspired by \citep{braverman2019multi} in the online learning domain, we first propose an impossibility theorem, proving that there is no $o(T)$-approximately incentive compatible mechanism that can guarantee a first-best user utility (formally defined later) asymptotically. Based on this, our objective is to propose a mechanism that guarantees a utility close to the second-best user utility.

Concretely, for any $\epsilon\in(0, \frac12)$, we propose a mechanism that is $O(T^{1-\epsilon}\log T)$-approximately incentive compatible. This mechanism consists of four phases: an exploration phase, during which a batch of ``test samples'' from each service provider are collected; an exploitation phase, where the best-performing provider is selected and asked to deliver the second-best user utility for the majority of queries throughout the mechanism; this is followed by two subsequent blind trust phases, which serve as incentives for providers to behave well in the first two phases. This mechanism guarantees a utility of $u_{SB}-O(T^{1-\epsilon}\log T+T^{2\epsilon})$, where $u_{SB}$ represents the second-best user utility.

\paragraph{Simulation experiments with real-world API settings.} Furthermore, we conduct experiments in simulated environments with real-world API price and performance settings to demonstrate the effectiveness of our mechanism. We designed a total of six strategies, including our own proposed strategy for providers. Through experimental evaluation, we demonstrate that our proposed strategy achieves the highest utility for the provider among all tested strategies, while also delivering considerable utility to the user.
\section{Related Work}

\paragraph{Auditing Large Language Models.} Numerous studies have investigated methods for auditing the disparities between the expected and actual outputs of commercial LLMs from various perspectives. \citet{chen2023chatgptsbehaviorchangingtime,10.1145/3630106.3659047} investigates how LLMs change their behavior over time. \citet{sun2025idiosyncrasieslargelanguagemodels} explores the idiosyncrasies present in the text generated by different language models. \citet{sun2025coincountinginvisiblereasoning} studies the auditing of invisible reasoning tokens in reasoning LLMs, specifically examining the existence of token count inflation with low-effort token injection. \citet{cai2025gettingpayforauditing} investigates the use of Trusted Execution Environments (TEEs), a hardware architecture, to provide cryptographic guarantees of model integrity. \citet{gao2025model} propose a concise and effective statistical testing scheme that quantifies changes in the output distribution due to quantization and watermarking by testing for the Maximum Mean Discrepancy between distributions. Recent works \citep{velasco2025llmoverchargingyoutokenization, velasco2025auditingpaypertokenlargelanguage} systematically explore how LLM providers misreport token usage to gain additional utility under identical string representations, as well as how to audit such behaviors.

\paragraph{Game Theory Meets Large Language Models.}The advancement of large language models brings new economic scenarios and problems. \citet{10.1145/3589334.3645511} introduce the token auction model, a mechanism design framework that allows multiple LLM agents to influence generated content through bids. \citet{bergemann2025economicslargelanguagemodels} provide economic frameworks for token pricing, and \citet{10.1145/3589334.3645366,sun2024mechanismdesignllmfinetuning} analyse the game-theoretic structure of fine-tuning and RLHF respectively. \citet{saig2024incentivizing} investigates how to incentivize LLM service providers to generate high-quality texts via cost-robust contracts.

\paragraph{Online Learning in Large Language Model Scenario.}
Algorithms from online learning can be utilized to control LLMs under uncertainty.
\citet{zhang2023banditrlhf} frame RLHF as a contextual bandit and use Thompson sampling to select the most informative comparison queries.
\citet{chen2024exploration} study the exploration--exploitation trade-off when an LLM generates candidate answers with an $\epsilon$-greedy wrapper.
\citet{wang2025auditingmab} treat safety auditing as an adversarial multi-armed bandit. Recent work   \citep{bouneffouf2025surveymultiarmedbanditsmeet} systematically explore future directions for combining multi-armed bandits with LLMs. Drawing inspiration from mechanism design in \citep{braverman2019multi}, we propose a novel model to address problems within the LLM scenario. This new context, in turn, presents new challenges for theoretical guarantees.
\section{The Model}
\label{sec:model}

\subsection{The User-Provider Delegation Game}

We model the interaction between the user and the service providers as a repeated Stackelberg game. The user acts as the principal, who commits to and announces a delegation mechanism at the beginning. The $K$ service providers act as agents who observe the mechanism and strategically respond to maximize their own utility over $T$ consecutive queries. Our goal is to design a mechanism in which the user is guaranteed to get the ``second-best'' service.

\paragraph{Model Components and Interaction Protocol.}

The game unfolds over $T$ discrete time units. In each time unit $t \in[T]$, the user has a query prompt $\omega_t$ sampled from a dataset $D$. The user selects one provider, denoted by $d_t \in\{1, \ldots, K\}$, to handle the query.

Each provider $i$ is characterized by a public price-per-token $p_i\in\sR_+$. Upon being selected, provider $d_t$ could secretly switch LLMs, leading to an actual cost-per-token\footnote{We will simplify it as \textit{cost} later when there is no ambiguity.} $c_t\in[0,p_{d_t}]$. This chosen cost leads to the generation of a cost-dependent output token sequence, $\tau_t$. In addition, the provider may report a different, potentially manipulated sequence, $\tau_t'$, to the user. All other providers $i \neq d_t$ do not take any action.

The user receives the reported sequence $\tau_t'$ and makes a payment of $p_{d_t} \cdot\left|\tau_t'\right|$. The user then evaluates the received response, obtaining a reward $v_t \in[0, R]$, where $R$ is a known upper bound of reward. The delegation choice becomes public knowledge and is added to the history of the game.

All possible output token sequences are assumed to have a length bounded by a constant $L$.

\begin{rem}
    For simplicity, we ignore the price and cost associated with processing the input prompt. As the prompt is user-provided, its length cannot be misreported, and any strategic cost control for processing it can be modeled similarly to the cost control for generating the output.
\end{rem}

\paragraph{Provider's Strategy Space and Truthful Behavior.} A dishonest provider can deviate from the \textit{truthful} behavior to increase its utility. We consider two main types of strategies:

\begin{enumerate}
    \item Cost control: For query $t$, the provider $d_t$ can secretly set $c_t\in[0, p_{d_t}]$ to a value other than the truthful one. For example, the provider could choose a cheaper, lower-quality model or sub-model to generate the response, or resort to quantization \citep{gong2024surveylowbitlargelanguage}.
    \item Token sequence reporting: For query $t$, the provider $d_t$ can report a token sequence $\tau'_t$ that is longer than the actual generated sequence $\tau_t$ to inflate its payment. This can be done by adding inconsequential reasoning tokens, or by using techniques like watermarking \citep{zhao2024sokwatermarkingaigeneratedcontent} to add tokens that are difficult for humans to discern. We assume such misreporting has negligible cost.\footnote{Note that how such inflation is realized is out of scope for this work.}
\end{enumerate}

In reality, the strategy space is discrete and finite (e.g., the provider can only choose from a limited number of models). In our theory, for mathematical convenience, \textbf{we assume a continuous strategy space}\footnote{In the discrete setting, our mechanism would introduce an additional approximation error of a single token cost (around $10^{-6}$ USD in practice) per query. It leads to a cumulative error of $O(T)$ over $T$ queries. Since the error per query is negligible, we adopt the continuous version in our model.}, where $c_t$ and the reported length $|\tau'_t|$ are both real numbers.

 Next, we define the truthful behavior for provider $d_t$ as follows:
\begin{enumerate}
    \item Truthful cost: The provider incurs a cost $c_t = p_{d_t}$.
    \item Truthful reporting: The provider reports the exact generated token sequence, i.e., $\tau_t' = \tau_t$.
\end{enumerate}

\paragraph{Objectives and Utilities.} The objectives of the user and providers are defined by their respective utility functions. The total utility for both the user and the provider is determined by the mechanism $M$ and strategies $(S_1,S_2,\dots,S_K)$ chosen.

The user aims to maximize the cumulative expected reward minus payments, with its expected utility defined as:

    \begin{equation*}
    U(M,S_1,S_2,\dots,S_K)=\mathbb E\left[\sum_{t=1}^T\left(v_t-p_{d_t} \cdot |\tau_t'|\right)\right].
    \end{equation*}

A strategic provider $i$ aims to maximize the cumulative expected difference between the truthful cost and the actual incurred cost, with its expected utility defined as:

    \begin{equation*}
    u_i^p(M,S_1,S_2,\dots,S_K) =\mathbb E\left[\sum_{t=1}^T\left(p_{i} \cdot|\tau_t'|-c_t \cdot |\tau_t|\right) \cdot \                       1_{d_t=i}\right].
    \end{equation*}

Note that if the provider behaves truthfully, its expected utility is 0. Thus intuitively, it directly represents the gains from a smart strategy.

The randomness in the expectation is sourced from the user's mechanism, the provider's strategy, the text generation process, or the data sampling.

\subsection{Goals of Mechanism Design}
\label{subsec:goals}

\paragraph{Dominant Strategy and Incentive Compatibility.} The user's objective is to design a mechanism that shapes provider behavior to protect the user's own utility. Without a well-designed mechanism, a dishonest provider would naturally adopt a highly exploitative strategy, such as incurring minimal cost while reporting maximum-length outputs.

The goal of the mechanism is not to force total truthfulness (we will prove it is impossible later), but to strategically incentivize providers to be truthful to a certain level. To measure the effectiveness of the mechanism, we need to introduce the concept of a \textit{dominant strategy}. An ideal mechanism ensures that each provider has an (approximately) optimal strategy, irrespective of the strategies chosen by other providers. This allows the proposed mechanism to produce a guaranteed outcome, as all providers will act according to the user's expectation. If such a mechanism is achievable, we call the mechanism is (approximately) incentive compatible.

\begin{defn}[Dominant Strategy]
Given a mechanism $M$, a strategy $S_i$ is a dominant strategy for provider $i$ if for any of its other strategies $S_i'$, and for any strategy

$S_{-i}=(S_1,\dots,S_{i-1},S_{i+1},\dots,S_K)$ of other providers, we have
\[
	u_i^p(M,S_i,S_{-i})\ge u_i^p(M,S_i',S_{-i}).
\]
\end{defn}

\begin{defn}[$o(T)$-Dominant Strategy]
Given a mechanism $M$, a strategy $S_i$ is a $o(T)$-dominant strategy for provider $i$ if for any of its other strategies $S_i'$, and for any strategy

$S_{-i}$ of other models, with a probability of $1-o\left(\frac1T\right)$, we have
\[
	u_i^p(M,S_i,S_{-i})\ge u_i^p(M,S_i',S_{-i})-o(T).
\]
\end{defn}

By this definition, we say a mechanism is $o(T)$-approximately incentive compatible if under this mechanism, an $o(T)$-approximately dominant strategy exists for every provider.

\paragraph{First-Best and Second-Best User Utility.} We can evaluate the performance of our mechanism against two clear benchmarks based on truthful provider behavior. As an optimal benchmark, the first-best user utility represents the user utility if the user could identify the single best provider and use them, assuming that the provider acts truthfully. It is defined as:

\begin{equation*}
u_{FB} = T\cdot\max_{i \in K} \left\{ \mu^r_i - p_i\mu^l_i\right\}.
\end{equation*}

where $\mu_i^r\in[0,R]$ and $\mu_i^l\in[1,L]$ is the provider's underlying expected reward and length of token sequence when incurring the truthful cost, respectively.

Assume provider $i^*$ can deliver $u_{FB}$. Similarly, we define the second-best user utility as follows:

\begin{equation*}
u_{SB} = T\cdot\max_{i \in K\setminus\{i^*\}} \left\{ \mu^r_i - p_i\mu^l_i \right\}.
\end{equation*}
\section{Second-Best Utility Guarantee Mechanism}
\label{sec:mech}

In this section, we propose our mechanism to mitigate the dishonest provider problem. There are four phases in our mechanism. We show the pseudocode of our mechanism in Mechanism~\ref{alg:random} and illustrate it in Figure~\ref{fig:mechanism}.

Informally, the optimal strategy under our mechanism (stated in Section~\ref{sec:thms}) is that the provider would be truthful in the exploration phase, give ``second-best'' service in the exploitation phase, and receive additional rewards in the blind trust phase. With this result in mind, we explain each phase below.

\begin{algorithm}
    \caption{Second-Best Utility Guarantee Mechanism}
    \label{alg:random}
    \begin{algorithmic}[1]
        \State \textbf{Input:} Total queries $T$, number of model providers $K$, max possible reward $R$, the price-per-token set $\{p_i\}$, $\epsilon\in(0,\frac12)$.
        \State $B \leftarrow T^{2\epsilon}$, $M \leftarrow T^{-\epsilon} \ln(KT)$
        \LineComment{Exploration phase starts}
        \For{each model provider $i = 1, \ldots, K$}
            \State Delegate $B$ queries to provider $i$.
            \State Calculate its average reward $\bar{v}_i$, average length of output token sequence $\overline{|\tau_i|}$, and average user utility $\bar u_i$.
            \State $\delta_i\leftarrow\frac{2\bar{v}_i}{p_iL} - \frac{2\overline{|\tau_i'|}}L$
        \EndFor

        \Statex
        \LineComment{Exploitation phase starts; break ties randomly}
        \State $i^* \leftarrow \arg\max_i \bar{u}_i, \bar u' \leftarrow \max_{i \neq i^*} \bar{u}_i$
        \State Inform provider $i^*$ of the value $\bar u'$.
        \State $T_{R} \leftarrow \left\lfloor T - \left(5K+\frac{2R}{\min_i\{p_i\}L}+\sum_{i\neq i^*}\delta_i\right)B-K\right\rfloor$
        \State \texttt{validated} $\leftarrow$ \textbf{true}
        \For{$j = 1, \dots, T_{R}$}
            \State Delegate 1 query to provider $i^*$.
            \If{$j>B$ \textbf{and} current average utility in this phase from $i^*$ is $<\bar u'-\frac{(b+p_{i^*}L)M}{3}$}
                \State \texttt{validated} $\leftarrow$ \textbf{false}
                \State \textbf{break}
            \EndIf
        \EndFor

        \Statex
        \LineComment{Blind trust phase I starts}
        \If{\texttt{validated}}
            \State Delegate $B$ queries to provider $i^*$.
        \EndIf
        \For{each service provider $i \neq i^*$}
            \State Delegate $B$ queries to provider $i$.
        \EndFor

        \LineComment{Blind trust phase II starts}
        \For{each service provider $i = 1, \ldots, K$}
            \State $\delta'_i \leftarrow  \delta_i + 3$
            \If{$\delta'_i \ge 0$}
                \State Delegate $\lfloor B\delta'_i\rfloor$ queries to provider $i$.
                \State With a probability $B\delta'_i - \lfloor B\delta'_i\rfloor$, delegate an additional query to provider $i$.
            \EndIf
        \EndFor
    \end{algorithmic}
\end{algorithm}

\paragraph{Exploration phase.} This phase is designed for the elicitation of providers' performance information. It operates by delegating each provider (expected to behave truthfully) on $B$ queries to estimate their performance. The best-performing provider is chosen for the next phase. The collected samples are also used to estimate the second-best utility.

\paragraph{Exploitation phase.} This is the key phase for the mechanism's user utility guarantee. It requires the best-performing provider to consistently deliver a user utility equivalent to the \textbf{second-best} among all providers from the exploration phase, for a duration of $T_R=O(T)$ queries. Recognizing the stochasticity in query samples and LLM generation, the performance check after each query involves a relax range of $O(T^{-\epsilon}\log T)$, which ensures that as long as the provider acts meeting the expectations, the probability of failing to complete the full $T_R$ queries is small ($o(\frac 1T)$).

\paragraph{Blind trust phase I} This phase primarily serves to reward the chosen provider for behavior that meets user expectations during the exploitation phase, while also compensating other providers. In this phase and the subsequent blind trust phase II, providers are permitted to engage in actions maximizing utilities, namely by incurring the minimum possible cost and reporting the maximum possible length of the token sequence. Note that, the exploitation phase requires an initial number of $B$ delegations to mitigate the effects of randomness. Therefore, without a reward of $B$-queries for the best provider, there is a clear incentive for it to deviate.

\paragraph{Blind trust phase II} As stated above, this phase is designed to incentivize honest behavior from all providers during the exploration phase. The number of queries is calculated to ensure that providers will achieve maximum utility when they incur truthful cost and report the truthful token sequence in the exploration phase.

\begin{figure*}[ht]
    \centering

    \includegraphics[width=\textwidth]{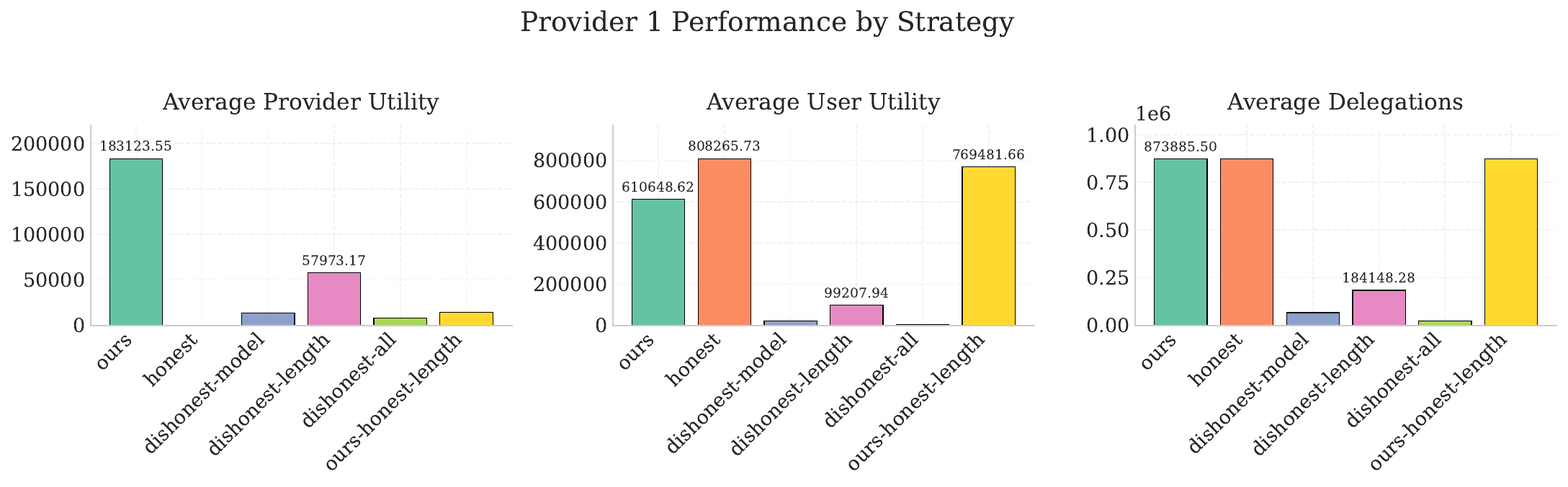}

    \caption{Simulation experiment where each provider enumerates strategies listed in Table \ref{tab:phase_strategy_mapping}. provider~1's expected provider utility, user utility, and number of delegations, averaged under permutations. Our proposed strategy gets the highest utility. The average delegations of \texttt{ours}, \texttt{honest}, \texttt{ours-honest-token} are the same.
}
    \label{fig:main}
\end{figure*}

\section{Our Results}
\label{sec:thms}

\begin{table*}[h!]
\centering
\caption{Overall provider strategies considered in our simulations. At each phase, the provider may choose a different strategy. Detailed descriptions of the strategy choices are provided in Table~\ref{tab:strategy_definitions}.
}
\begin{tabular}{lcccc}
\toprule
\textbf{Provider Strategy} & \textbf{Exploration Phase} & \textbf{Exploitation Phase} & \textbf{Blind Trust Phase I} & \textbf{Blind Trust Phase II} \\
\midrule
Ours & Honest & Second-best & Worst & Worst \\
Honest & Honest & Honest & Honest & Honest \\
Dishonest-model & Honest & Worst model & Worst & Worst \\
Dishonest-length & Honest & Worst length & Worst & Worst \\
Dishonest-all & Worst & Worst & Worst & Worst \\
Ours-honest-length & Honest & Second-best with honest length & Worst & Worst \\
\bottomrule
\end{tabular}
\label{tab:phase_strategy_mapping}
\end{table*}

In this section, we present the main theoretical results. Firstly, we show that no $o(T)$-approximate incentive compatible mechanism can achieve the first-best user utility asymptotically. Consequently, we demonstrate that, given the assumption of certain model capabilities, the proposed mechanism is $O(T^{1-\epsilon}\log T)$-approximately incentive compatible; that is, it is an $O(T^{1-\epsilon}\log T)$-dominant strategy for all service providers. Concurrently, we prove that this mechanism yields a user utility close to the second-best user utility $u_{SB}$.

We provide theoretical guarantees for the continuous strategy space. \textbf{The complete proofs are detailed in Appendix~\ref{sec:app_proofs}}.

\subsection{Notations}

To formally link a provider's actions to outcomes, we model the relationship between cost, output length, and reward.

For a random query $\omega \sim D$, provider $i$'s incurring cost $c$ induces a distribution $\mathcal T_{i,c}$ over the generated token sequence, denoted by the random variable $\tau_{i,c}$, and a distribution $\mathcal V_{i,c}$ over the user's reward, denoted by the random variable $v_{i,c}$. The distribution of the sequence length $|\tau_{i,c}|$ is consequently induced by $\mathcal T_{i,c}$.

We assume providers have an internal belief about their cost-performance trade-offs, modeled by two functions, $g_i(c)$ and $h_i(c)$, where $c\in[0,p_i]$, under the boundary condition $g_i(p_i)=\mu_i^l$ and $h_i(p_i)=\mu_i^r$ (both $\mu_i^l$ and $\mu_i^r$ are defined in Section~\ref{sec:model}). The expected length is defined such that $\E[|\tau_{i,c}|] = g_i(c)$. The expected reward is defined such that $\E[v_{i,c}] = h_i(c)$. We assume they are Lipschitz continuous.

\subsection{Main Results}

First, we state the impossibility of designing an $o(T)$-approximate incentive compatible mechanism that allows a user to achieve the first-best utility.

\begin{restatable}[The impossibility of the first-best user utility]{thm}{thmimp}
    \label{thm:impossibility}
    For any $\xi\in(0,1)$, there is no $o(T)$-approximately incentive compatible mechanism can guarantee an expected user utility of $\xi u_{FB}+(1-\xi)u_{SB}$, where $u_{FB}$ and $u_{SB}$ are the first-best and the second-best user utility (defined in Section~\ref{subsec:goals}).
\end{restatable}

\textit{Proof sketch:} We prove this theorem by contradiction. We construct a sufficiently large number of type sets (formally defined in Appendix~\ref{sec:app_proofs}), among which the utility from the best provider in each type set exhibits certain variance, while the utility from the second-best providers remains constant. We derive that if a mechanism were to exist that could guarantee an expected utility of $\xi u_{FB}+(1-\xi)u_{SB}$, it would necessitate the user to delegate tasks to the best provider more than $T$ queries, which is an impossibility.

Next, we state the first positive result for our mechanism.

\begin{restatable}[$O(T^{1-\epsilon}\log T)$-dominant strategy of all model providers]{thm}{thmpre}
    \label{thm:pre}
    If for all providers $i\in[K]$ and choices of cost $c\in [0, p_i]$, the cost-performance functions $h_i(c), g_i(c)$ satisfies:

    \begin{itemize}
        \item $\frac{\dif h_i}{\dif c}(c)-p_i\frac{\dif g_i}{\dif c}(c)\ge\gamma$ \text{for some $\gamma>0$},
        \vspace{1ex}
        \item $\frac{\dif h_i}{\dif c}(c)-c\frac{\dif g_i}{\dif c}(c)-g_i(c)\ge0$,
    \end{itemize}

    then the following strategy is an $O(T^{1-\epsilon}\log T)$-dominant strategy for provider $i$ in Mechanism~\ref{alg:random}:
    \begin{itemize}
        \item In the exploration phase, incur the truthful cost and report the truthful output token sequence for each query;
        \item If $i=i^*$, in the exploitation phase, for each query, calculate a pair $(c',l'))$ s.t.

        \begin{equation*}
            (c',l')=\argmax_{\substack{c\in[0,p_i], l\in[g_i(c),L]\\ h_i(c)-p_il\ge\bar u'}}\{p_il-cl\},
        \end{equation*}

        where $i^*, \bar u'$ are chosen by the previous phase.
        Then incur a cost of $c'$ and generate a corresponding token sequence $\tau$. If $|\tau|\le l'$, report an output token sequence of length $l'$; otherwise, report the truthful token sequence.

        \item In the blind trust phases, incur a zero cost and report an output token sequence of length $L$ for each query.
    \end{itemize}
    Note that $p_i$ is the price-per-token and the truthful cost-per-token of provider $i$, $\mu_i^r$ and $\mu_i^l$ are expected reward and length of output token sequence when incurring the truthful cost, and $L$ is the max possible length of token sequence.
 \end{restatable}

\textit{Proof sketch:} The main idea of the proof is backward induction. For the two blind trust phases, dishonest providers can do their worst since there is no limitation for them.

For the exploitation phase, we prove that the best provider can obtain an $O(T^{1-\epsilon}\log T)$-approximate utility by adopting a best choice with respect to the expected second-best user utility constraint. We establish the proof by constructing an ideal optimal strategy and demonstrating that the provider utility of our proposed strategy deviates from that of the optimal strategy by at most $O(R M)=O\left(T^{1-\epsilon} \log T\right)$.

For the exploration phase, we prove that, given the constraint of the blind trust phase, we prove that for any provider, incurring the truthful cost and reporting the truthful token sequence is a dominant strategy, by showing that the total provider utility is maximized under this strategy.

This theorem relies on two assumptions. First, we assume a positive correlation between a provider's incurred cost and the utility delivered to the user. This assumption is justifiable within our context, i.e., for each provider, in their own choices, LLMs of higher cost always yield better performance. The second assumption is that, as cost-per-token rises, the growth in reward to the user must outpace the increase in total cost.

By this theorem, we know that our mechanism is $O(T^{1-\epsilon}\log T)$-approximately incentive compatible.

\begin{cor}
    Mechanism~\ref{alg:random} is $O(T^{1-\epsilon}\log T)$-approximately incentive compatible.
\end{cor}

Finally, we prove that our mechanism guarantees a utility asymptotically close to the second-best user utility. The proof of this theorem is straightforward, as the user utility in our mechanism is primarily derived from the $T_R$ queries during the exploitation phase.

\begin{restatable}[Second-best user utility guarantee]{thm}{thmuti}
    \label{thm:uti}
    If all providers act according to Theorem~\ref{thm:pre}, the user could get an expected utility of at least $u_{SB}-O(T^{1-\epsilon}\log T+T^{2\epsilon})$.
\end{restatable}
\section{Experiments}
\label{sec:exp}

\begin{figure*}[t]
    \centering

    \includegraphics[width=\textwidth]{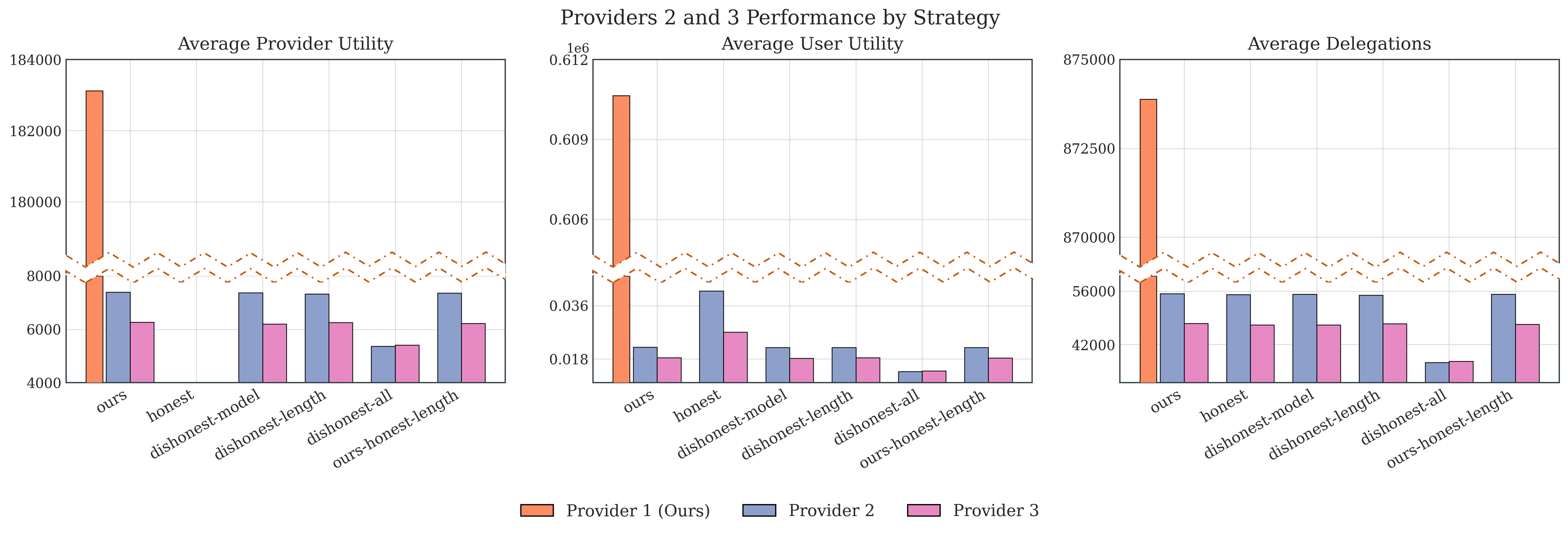}

    \caption{Average provider utility, user utility (from each provider), and number of delegations for provider~2 and provider~3 adopting different strategies when provider~1 adopts our proposed strategy. Since provider~1 has the best-performing LLM, they can not achieve a decent utility.
}

    \label{fig:p23}
\end{figure*}

To demonstrate the efficacy of our mechanism, we conduct simulation experiments. This section presents the experiments in detail.

Although our theoretical analysis is confined to the study of continuous action spaces, we conduct our experiments in a scenario where the action spaces for both cost control and token sequence length are discrete to demonstrate the practical effectiveness of our mechanism. Analogous to Theorem~\ref{thm:pre}, in this context, we employ the upper difference as a proxy for the differential to construct a discrete version of assumptions  (we omit the detailed formulation of this assumption due to lack of space). The proposed strategy in Theorem~\ref{thm:pre} remains the same in the subsequent experiments, with slight modifications for a discrete action space.

\subsection{Setup}

\paragraph{Simulation Scenarios.} Our scenario involves three independent providers, each employing three models. We use the price of the best model as the provider's price, and the price of each model as the provider's different cost options. When a user delegates a query to a provider, the provider selects a model according to its strategy, returns the output token sequence to the user, and reports the number of tokens used (which may be inaccurate), thereby charging the user. In addition to our proposed strategy, we consider several other strategies, as shown in Table~\ref{tab:phase_strategy_mapping}. Provider~1 employs three models: gpt-5-high~\citep{gpt-5}, deepseek-r1~\citep{deepseekai2025deepseekr1incentivizingreasoningcapability}, and gpt-5-medium. Provider~2 employs o3-mini~\citep{o3mini}, o1-mini~\citep{o12024}, and deepseek-r1. Provider~3 offers claude-4.0~\citep{claude}, o1-mini, and deepseek-r1. All model pricing information is obtained from the official websites of the respective models.

\paragraph{Dataset.}  We select the maximum flow task from NLGraph~\citep{wang2023can} as the dataset for our experiments for the following reasons: (1) the maximum flow task enables the generation of a large number of experimental samples; (2) current LLMs' performance on the task has not yet reached saturation. Following~\citet{wang2023can}, we use the ``partial credit'' as the evaluation metric for a single query (which is also the user's reward), ranging from 0 to 1, with higher values indicating better performance.

\paragraph{Hyperparameters.} In our main experiment, we set $T$ to 1 million. To reduce costs and accelerate the process, we run 2,000 results for each LLM on NLGraph and repeatedly sample these 2,000 results during the experiment. We set $\epsilon$ to 0.3, and $L$ to 38,058 (the maximum output length in our samples). To verify the user utility guarantee, we also conduct a supplementary experiment by varying $T$ from 1 million to 2 million.

\textbf{We verify that the assumptions discussed at the beginning of this section are satisfied. And for this task, provider 1 has the best-performing LLM (gpt-5-high)}.

\begin{table*}[h!]
\centering
\caption{Description of strategy choices considered in our simulations.}
\begin{tabular}{lcc}
\toprule
\textbf{Strategy Choice} & \textbf{LLM Used}  & \textbf{Output Length Reported} \\
\midrule
Honest & Largest cost &  Truthfully \\
Second-best with honest length & Same as Theorem~\ref{thm:pre} &  Truthfully \\
Second-best & Same as Theorem~\ref{thm:pre} &  Same as Theorem~\ref{thm:pre} \\
Worst model & Lowest cost &   Truthfully \\
Worst length & Largest cost &  Report the maximum limit \\
Worst & Lowest cost &  Report the maximum limit \\
\bottomrule
\end{tabular}
\label{tab:strategy_definitions}
\end{table*}

\subsection{Results}

\begin{figure}[t]
    \centering

    \includegraphics[width=0.4\textwidth]{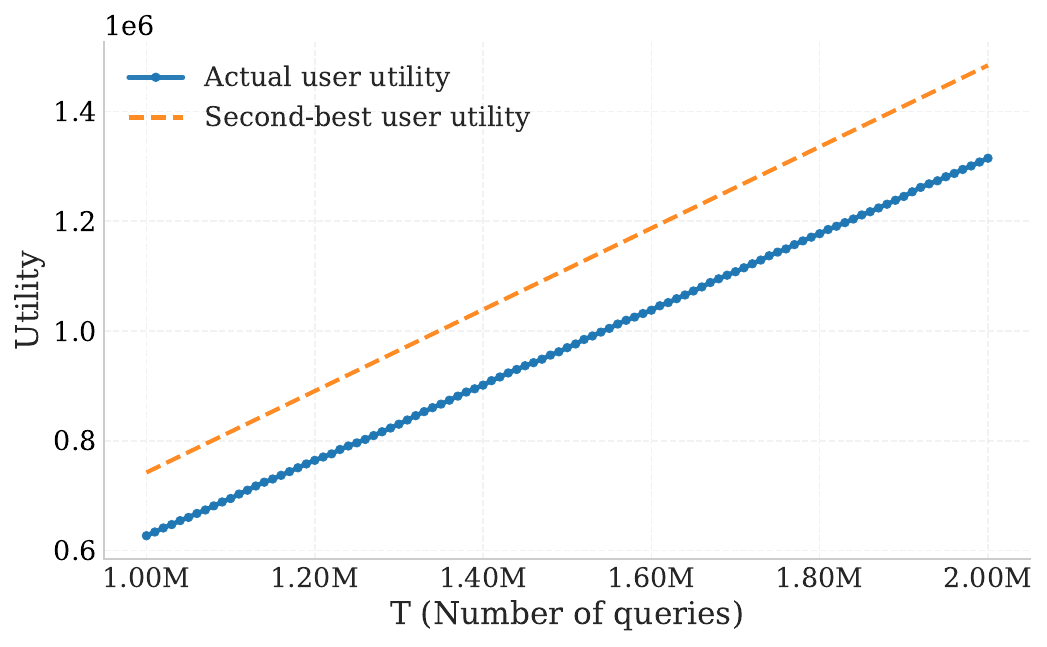}
    \caption{User utility of provider~1 as $T$ increases from 1 million to 2 million. We also plot a theoretical reference, $u_{SB}$ (defined in Section~\ref{sec:model}).}
    \label{fig:t}
\end{figure}

In our experimental evaluation, we compare several different strategies (Table~\ref{tab:phase_strategy_mapping}) to validate the efficacy of our proposed mechanism.

As a baseline for comparison, the \texttt{honest} strategy consistently reports the model that is optimal for the user (based on the assumptions in Theorem~\ref{thm:pre}, a higher cost correlates with greater user utility). Consequently, the provider utility for this strategy is perpetually zero. In contrast, the \texttt{dishonest-all} strategy perpetually employs the worst LLM and bills for the maximum possible token length.

We further consider two more sophisticated dishonest strategies, \texttt{dishonest-token} and \texttt{dishonest-length}. These strategies behave honestly during the exploration phase and attempt to be selected for the subsequent exploitation phase. However, during the exploitation phase, they resort to deploying the worst LLM or billing for the maximum possible token length, respectively. When successful, these strategies could severely degrade user utility.

Furthermore, to demonstrate that misreporting the token sequence length is necessary for a strategic provider, we implement an additional strategy, \texttt{ours-honest-length}. This variant  only attempts to substitute the model while reporting the truthful output length for billing.

\paragraph{Metrics.} For our main experiment, we calculate the average provider utility, user utility, and number of delegations for different strategy permutations ($6^3$ combinations in total). For provider~1 (Figure~\ref{fig:main}), we compute the average results under different strategy permutations adopted by provider~2 and provider~3. For provider~2 and provider~3 (Figure~\ref{fig:p23}), we compute the average results under different strategy permutations when provider~1 adopts our proposed strategy in Theorem~\ref{thm:pre}. For the supplementary experiment (Figure~\ref{fig:t}), we consider the situation that all providers follow our proposed strategy, as stated in Theorem~\ref{thm:uti}.

\paragraph{Optimality of Our Proposed Strategy.} We show our main experiment result in Figure~\ref{fig:main} and Figure~\ref{fig:p23}. Under all permutations of provider strategies, adhering to our proposed strategy yields the optimal average provider utility for the highest-performing provider (provider 1), while concurrently generating substantial user utility. Conversely, when provider 1 adopts our proposed strategy, providers 2 and 3 are precluded from being selected by the user during the exploitation phase, regardless of their chosen strategies. This outcome validates the robustness of our mechanism.

\paragraph{User Utility Guarantee.} The results of our supplementary experiment are presented in Figure~\ref{fig:t}. As illustrated in the figure, the user utility exhibits a linear increase with $T$ and close to the second-best user utility, which is consistent with the statement in Theorem~\ref{thm:uti}. This demonstrates that our mechanism maintains stable effectiveness across different values of $T$.

\section{Limitation}

Our work has the following limitations:

First, our exploration of the model is limited. Although we have established a novel game model, there is considerable room for further investigation. Our theory does not delve into the mechanism design for scenarios where the token sequence length and the action space for cost control are discrete. Furthermore, our proposed mechanism requires all providers to have complete prior knowledge of the model's capabilities.

Second, our model simplifies several real-world complexities. We assume no collusion between providers; however, drawing a parallel to auction theory, the first-best and second-best providers may have strong incentives to engage in collusive behavior if mutual recognition exists. Additionally, we assume that the user (mechanism designer) acts truthfully within the mechanism; relaxing this assumption could undermine the mechanism. For example, a user might be incentivized to strategically misreport $B$, the number of queries per provider in the exploration phase.
\vspace{-1mm}
\section{Conclusion and Future Work}
\label{sec:con}
This work confronts the critical issue of service provider dishonesty within the black-box LLM API market, by shifting the focus from technical detection to the lens of algorithmic game theory. We formalize this challenge as a user-provider delegation game, for which we establish a foundational impossibility theorem. In response, we propose an approximately incentive-compatible mechanism that guarantees the user a near-optimal utility, the effectiveness of which is validated through comprehensive simulations grounded in real-world API parameters. Our mechanism can be effectively implemented in aggregated API platforms and enterprise clients, both of which have the leverage to contract with different providers.

To the best of our knowledge, this work is the first to propose a game-theoretic framework towards ensuring a more transparent and trustworthy market for large language model services. Future research could extend our model to encompass multiple users, fostering a more competitive and fair market environment. Another valuable problem is to assume the user has a budget constraint, which is a more realistic assumption and a challenge for our model. By establishing a framework grounded in economic principles, we hope to inspire further research at the intersection of artificial intelligence and mechanism design, fostering a healthier and more reliable ecosystem for all users of large language models.

\vspace{-2mm}

\begin{acks}
We would like to express our sincere gratitude to Kaifeng Lyu for providing valuable feedback on a preliminary version of this manuscript. We are also thankful to Minrui Luo for carefully reading of our proofs and for providing several useful suggestions. Additionally, we would like to thank Yu Chen, Zhaohua Chen, Yaoxin Ge, Zixin Gu, Xuanyu Li and Dengji Zhao for their helpful discussions in the early stages of this work. Any remaining errors are our own.
\end{acks}

\vspace{-1mm}

\bibliographystyle{ACM-Reference-Format}
\balance
\bibliography{ref}

\appendix
\section{Proofs}

\label{sec:app_proofs}

\subsection{Impossibility of First-Best User Utility}

For clarity, we first provide definitions of types and type sets.

\begin{defn}[Types and Type Sets]
    For any service provider $i$, we say its type consists of all private information (e.g. $\mu_i^r,u_i^p$ and $g_i$ defined in Section~\ref{sec:model} and Section~\ref{sec:thms}),. A type set is the set of types for all providers.
\end{defn}

Note that although the price-per-token $p_i$ for provider $i$ is public knowledge, the utility function is private. For simplicity, we will also use $p_i$ in the proof without further explanation.

\thmimp*

\begin{proof}
    We prove by contradiction.

    Assume there exists an $o(T)$-approximate incentive-compatible mechanism that guarantees the user a worst-case utility of $\xi u_{FB}+(1-\xi)u_{SB}$. Denote $i^*$ as the provider who delivers $u_{FB}$, and $i'$ as the provider who delivers $u_{SB}$. We construct $N>\exp\left(\frac{1}{\xi-\lambda}\right)$ type sets, where $\lambda=\frac{\xi}2$, and for the $j$-th type set, we set $u_{FB}=T\cdot (\frac12+\frac j{2N})$ and $u_{SB}=\frac T2$. we denote $u^{(j)}=\frac12+\frac j{2N}$ and $u^{\circ}=\frac12$. Within these type sets, we consistently use the notation $(i^*,i')$. The types of all other service providers are identical across all type sets. Let $F_j(c)=\frac{cu^{(j)}}{p_{i^*}}$, where $p_{i^*}$ keeps unchanged for all type sets.

    For the $j$-th type set, let $x_jT$ denote the expected number of delegations to provider $i^*$ under the assumed mechanism, and let $y_jT$ be the expected utility to the user. The total expected cost incurred by provider $i^*$ could be represented by $F^{-1}_j(y_j)T$.

    Because the mechanism is $o(T)$-approximately incentive compatible, all service providers will adopt their $o(T)$-approximate dominant strategies. During the mechanism running on the $j$-th type set, any provider $k$ will prefer their dominant strategy for the $j$-th type set over the dominant strategy for any other type set $j'$ (where $j\neq j'$), i.e., $x_jp_{i^*}T-F^{-1}_j(y_j)T\ge x_{j'}p_{i^*}T-F^{-1}_j(y_{j'})T-o(T)$. Hence, when $T$ is sufficiently large, we have

\begin{equation}\label{ineq:j_is_best}
\begin{aligned}
    x_jp_{i^*}T-F^{-1}_j(y_j)T&> x_{j'}p_{i^*}T-F^{-1}_j(y_j')T-\frac{\lambda p_{i^*}T}{u^{(j)}N}\\
    \Leftrightarrow x_jp_{i^*}T-\frac{y_jp_{i^*}T}{u^{(j)}}&> x_{j'}p_{i^*}T-\frac{y_{j'}p_{i^*}T}{u^{(j)}}-\frac{\lambda p_{i^*}T}{u^{(j)}N}\\
    \Leftrightarrow x_ju^{(j)}-y_j&> x_{j'}u^{(j)}-y_{j'}-\frac{\lambda}N.\\
\end{aligned}
\end{equation}

Note that we only consider $T$ to be non-constant in asymptotic notations. On the other hand, since the expected utility of $i^*$ is not less than $0$ (they can achieve it by incurring $0$ cost at all times), we have

\begin{equation*}
    \begin{aligned}
        x_jp_{i^*}T-F^{-1}(y_j)T&\ge0\\
        \Rightarrow x_ju^{(j)}-y_j&\ge0.
    \end{aligned}
\end{equation*}

Therefore, by enumerating all $j$ for inequality~(\ref{ineq:j_is_best}), we know that

\begin{equation*}
\begin{cases}
    &y_1\le x_1u^{(1)}\\
    &y_2\le u^{(2)}(x_2-x_1)+y_1+\frac{\lambda}N\\
    &y_3\le u^{(3)}(x_3-x_2)+y_2+\frac{\lambda}N\\
    &\dots\\
    &y_N\le u^{(N)}(x_N-x_{N-1})+y_{N-1}+\frac{\lambda}N.
\end{cases}
\end{equation*}

Thus for all $j\in[N]$ we have

\begin{equation}\label{ineq:y_i}
    \begin{aligned}
    y_j&\le u^{(j)}x_j+x_{j-1}(u^{(j-1)}-u^{(j)})+x_{j-2}(u^{(j-2)}-u^{(j-1)})\\
    &+\dots+x_1(u^{(1)}-u^{(2)})+\frac{j\lambda}N\\
    &=u^{(j)}x_j+\sum_{k=1}^{j-1}x_{k}(u^{(k)}-u^{(k+1)})+\frac{j\lambda}N.
    \end{aligned}
\end{equation}

Consider an upper bound for the user's utility, that is, apart from delegating provider $i^*$ for $x_jT$ queries, the user delegates provider $i'$ for the remaining $(1-x_j)T$ queries (as an upper bound). For provider $i'$, it is impossible to bring more expected utility to the user, because they need to incur the truthful cost and report the truthful token sequences to bring such an expected utility to the user, and hence their own expected utility is already $0$. At this point, the upper bound of the user's utility is $\left[y_j+(1-x_j)u^{\circ}\right]T$. Therefore, $\left[y_j+(1-x_j)u^{\circ}\right]T\ge(\xi u^{(j)}+(1-\xi)u^{\circ})T$.

By substituting $y_j$ from the inequality~(\ref{ineq:y_i}), we know that

\begin{equation*}
    \begin{aligned}
        &\left[u^{(j)}x_j+\sum_{k=1}^{j-1}x_{k}(u^{(k)}-u^{(k+1)})+(1-x_j)u^{\circ}+\frac{j\lambda}N\right]T\\
        \ge&(\xi u^{(j)}+(1-\xi)u^{\circ})T\\
        \Rightarrow&(u^{(j)}-u^{\circ})x_j\ge\xi(u^{(j)}-u^{\circ})+\sum_{k=1}^{j-1}x_{k}(u^{(k+1)}-u^{(k)})-\frac{j\lambda}N\\
        \Rightarrow&x_j\ge\xi+\sum_{k=1}^{j-1}\frac{x_{k}(u^{(k+1)}-u^{(k)})}{(u^{(j)}-u^{\circ})}-\lambda=\xi+\sum_{k=1}^{j-1}\frac{x_k}j-\frac{j\lambda}N.\\
    \end{aligned}
\end{equation*}

Next we use induction to prove $x_j\ge(\xi-\frac{j\lambda}N)\sum_{k=1}^j\frac1k$.

When $j=1$, $x_1=\xi-\frac{\lambda}N$. Assume (by induction) for $j^\circ\in[N]$, we have $x_{j}\ge(\xi-\frac{j\lambda}N)\sum_{k=1}^{j}\frac1k$ holds for all $1\le j\le j^\circ$. For $j^\circ+1$, we know that

\begin{equation*}
    \begin{aligned}
        x_j&\ge\xi+\sum_{k=1}^{j-1}\frac{x_k}j-\frac{j\lambda}N\\
        &\ge\xi-\frac{j\lambda}N+\sum_{k=1}^{j-1}\frac{\left(\xi-\frac{k\lambda}{N}\right)\sum_{l=1}^{k}\frac 1l}j\\
        &\ge\left(\xi-\frac{j\lambda}{N}\right)\left(1+\frac1j\sum_{k=1}^{j-1}\sum_{l=1}^{k}\frac 1l\right)\\
         &=\left(\xi-\frac{j\lambda}{N}\right)\left(1+\sum_{l=1}^{j-1}\frac{j-l}{jl}\right)\\
          &=\left(\xi-\frac{j\lambda}{N}\right)\sum_{l=1}^j\frac1l.
    \end{aligned}
\end{equation*}

Therefore, $x_j\ge\left(\xi-\frac{j\lambda}{N}\right)\sum_{k=1}^j\frac1k>\left(\xi-\frac{j\lambda}{N}\right)\ln j$ for all $j\in[N]$ and we have

\begin{equation*}
    \begin{aligned}
        x_N&>(\xi-\lambda)\ln N\\
        &=(\xi-\lambda)\ln\exp\left(\frac{1}{\xi-\lambda}\right)=1.
    \end{aligned}
\end{equation*}

It implies that the user delegates $x_NT>T$ queries for $i^*$, which is a contradiction.

\end{proof}

\subsection{Properties of Our Mechanism}

We first state two lemmas used in the following proofs.

\begin{restatable}[]{lemma}{lemtime}
    \label{lem:time}
    The total completed queries of mechanism \ref{alg:random} are at most $T$.
\end{restatable}

We omit this proof due to lack of space.

\begin{restatable}[]{lemma}{lemgap}
    \label{lem:gap}
    For all $i\in[K]$, if the cost-performance functions $h_i(c), g_i(c)$ satisfies that there exists a $\gamma>0$ s.t. $\frac{\dif h_i}{\dif c}(c)-p_i\frac{\dif g_i}{\dif c}(c)\ge\gamma$, then for all pairs $(c',l'),(c',l'')$ s.t.
    \begin{equation*}
        \begin{aligned}
    (c',l')=\argmax_{\substack{c\in[0,p_i], l\in[g_i(c),L]\\ h_i(c)-p_il\ge\bar u'}}\{p_il-cl\},\\(c'',l'')=\argmax_{\substack{c\in[0,p_{i}], l\in[g_i(c),L]\\ h_{i}(c)-p_{i}l\ge \bar u'-\frac{(b+p_{i}L)M}3}}\{p_{i}l-cl\},
        \end{aligned}
    \end{equation*}
    we have $p_i(l''-l')-c''l''+c'l'=O(M)$.
\end{restatable}

\begin{proof}
    Let $H(c, l)=p_il-cl$, $\Delta_M=\frac{(b+p_{i}L)M}3$, and $P_1,P_2$ denote two programs in the statement respectively.

    $(c'',l'')$ may not be feasible for $P_1$, as the value of $h_i\left(c^{\prime \prime}\right)-p_i l^{\prime \prime}$ could be in the interval $\left[\bar{u}^{\prime}-\Delta_M, \bar{u}^{\prime}\right)$. To construct a feasible solution for $P_1$, we need to increase the value of the expression $h_i(c)-p_i l$ by at most $\Delta_M$.

    By the assumption

$$
\frac{\dif h_i}{\dif c}(c)-p_i\frac{\dif g_i}{\dif c}(c) \ge\gamma>0,
$$
we know that the user utility function $\phi(c)=h_i(c)-p_i g_i(c)$ is strictly increasing in $c$ at a rate of at least $\gamma$. We will leverage this property to construct a feasible solution for $P_1$ starting from $\left(c^{\prime \prime}, l^{\prime \prime}\right)$.

We construct a new point $(\tilde{c}, \tilde{l})$ by perturbing $c^{\prime \prime}$ by a small positive amount $\Delta_c$. Let $\tilde{c}=c^{\prime \prime}+\Delta_c$. To ensure feasibility while controlling the change, we set $\tilde{l}$ to its lower bound (constrained by both $P_1$ and $P_2$), i.e., $\tilde{l}=g_i(\tilde{c}) $.

We know that $l^{\prime \prime} \geq g_i\left(c^{\prime \prime}\right) $. Therefore:

$$
\phi\left(c^{\prime \prime}\right)=h_i\left(c^{\prime \prime}\right)-p_i g_i\left(c^{\prime \prime}\right)  \geq h_i\left(c^{\prime \prime}\right)-p_i l^{\prime \prime} \geq \bar{u}^{\prime}-\Delta_M.
$$

We want to choose $\Delta_c$ such that $\phi(\tilde{c})=\phi\left(c^{\prime \prime}+\Delta_c\right) \geq \bar{u}^{\prime}$. This requires an increase in the value of $\phi$ by at most $\Delta_M$.
The change in $\phi$ is:

\begin{equation*}
    \begin{aligned}
       \phi\left(c^{\prime \prime}+\Delta_c\right)-\phi\left(c^{\prime \prime}\right)&=\int_{c^{\prime \prime}}^{c^{\prime \prime}+\Delta_c}\left(\frac{\dif h_i}{\dif c}(c)-\frac{\dif g_i}{\dif c}(c) p_i \right) \dif c \\
       &\geq \int_{c^{\prime \prime}}^{c^{\prime \prime}+\Delta_c} \gamma \dif c=\gamma \Delta_c.
    \end{aligned}
\end{equation*}

To achieve the required increase, we set $\gamma \Delta_c=\Delta_M$, which gives $\Delta_c=\frac{\Delta_M}{\gamma}$.
Since $\Delta_M=O(M)$ and $\gamma$ is a positive constant, we have $\Delta_c=O(M)$.

We have now constructed a point $(\tilde{c}, \tilde{l})=\left(c^{\prime \prime}+\Delta_c, g_i\left(c^{\prime \prime}+\Delta_c\right) \right)$ that is feasible for $P_1$. The change in the $c$ coordinate is $\left|\tilde{c}-c^{\prime \prime}\right|=\Delta_c=O(M)$.
Since $g_i$ is Lipschitz continuous, the change in the $l$ coordinate is also bounded:

$$
\left|\tilde{l}-g_i\left(c^{\prime \prime}\right) \right|=\left|g_i\left(c^{\prime \prime}+\Delta_c\right) -g_i\left(c^{\prime \prime}\right) \right| \leq \kappa \Delta_c \mu_i^l=O(M),
$$

where $\kappa$ is the Lipschitz constant of $g_i$. The distance between $\left(c^{\prime \prime}, l^{\prime \prime}\right)$ and $(\tilde{c}, \tilde{l})$ is thus of order $O(M)$.

On the other hand, the objective function $H(c, l)=p_il-cl$ has bounded partial derivatives on the compact feasible domain, which implies it is also Lipschitz continuous. This property guarantees that a perturbation of order $O(M)$ in the input variables will result in a change of at most order $O(M)$ in the function's value, i.e.,

\begin{equation*}
    \begin{aligned}
H\left(c^{\prime \prime}, l^{\prime \prime}\right)-H(\tilde{c}, \tilde{l})&=O\left(\left\|\left(c^{\prime \prime}, l^{\prime \prime}\right)-(\tilde{c}, \tilde{l})\right\|_{\infty}\right)\\
&=O(M).
    \end{aligned}
\end{equation*}

By construction, ($\tilde{c}, \tilde{l}$) is a feasible solution for $P_1$. By the definition of optimality, $H(c',l')$ must be greater than or equal to the value of the objective function at any feasible point, i.e., $H(c',l')\geq H(\tilde{c}, \tilde{l})$.
Combining our results, we get:

\begin{equation*}
    \begin{aligned}
p_i(l''-l')-c''l''+c'l'&=H(c'',l'')-H(c',l')\\
    &\leq H(c'',l'')-H(\tilde{c}, \tilde{l})\\
    &=O(M).
    \end{aligned}
\end{equation*}
\end{proof}

Next we give the proof of Theorem~\ref{thm:pre}, we do not repeat it here due to lack of space.

\begin{proof}[\textbf{Proof of Theorem~\ref{thm:pre}}]

    We prove that the strategy of any model provider at each step is approximately optimal under any historical conditions, using backward induction, a common method in sequential games.

First, we begin the proof with the analysis of the blind trust phases. Regardless of the previous delegation history, any action taken by a provider during this phase has no impact on the number of future delegations they or any other provider will receive. Therefore, in blind trust phases, a service provider will choose to incur a minimum cost, and misreport a token sequence with a length of $L$, which is optimal under any circumstances (i.e., a dominant strategy).

Next, we analyze the exploitation phase (with consideration of the blind trust phase I). It is important to note that in a sequential game, if all players are sufficiently rational, they will analyze the subsequent game situation after making a current choice and select the optimal decision for the entire game. We only need to consider the case where $i=i^*$, as when $i\neq i^*$, they have no room for action and receive the maximized utility by incurring a zero cost, and misreporting a token sequence with a length of $L$ (which has been analyzed above).

Let $F(c,l)=h_{i^*}(c)-p_{i^*}l$ be the expected user utility when provider $i^*$ incurs a cost of $c$ and report the token sequence with a length of $l$. The history from the exploration phase leads to two cases: either $F(p_{i^*}, \mu_{i^*}^l)=\mu_{i^*}^r-p_{i^*}\mu_{i^*}^l<\bar u'$ or $F(p_{i^*}, \mu_{i^*}^l)\ge \bar u'$. If $F(p_{i^*}, \mu_{i^*}^l)<\bar u'$, the provider $i^*$ should immediately incurs a zero cost and misreports a token sequence of length $L$ (deviating), because they cannot obtain an expected utility greater than always deviating throughout the exploitation phase and the blind trust phase I.

If $F(p_{i^*}, \mu_{i^*}^l)\ge \bar u'$, let $\bar v_{t}$, $\overline{|\tau_{t}|}$ and $\overline{|\tau_{t}'|}$ be the average reward, average length of truthful and reported token sequences delivered by provider $i^*$ after $t$ time units in the exploitation phase. When $KT$ is sufficiently large, by Hoeffding's inequality, we know that

\begin{equation*}
\begin{aligned}
\operatorname{P}\left[\left|\bar{v}_t-h_{i^*}(c')\right| \geq \frac{bM}{3}\right] &\leq 2 \exp \left(-\frac {2B\left(\frac{bM}{3}\right)^2}{b^2}\right)\\
&= 2\exp\left(-\frac{2(\ln KT)^2}{9}\right)\\
&\le 2\exp\left(-\frac{18\ln KT}{9}\right)\\
&\leq \frac 2{(KT)^2}.
\end{aligned}
\end{equation*}

Therefore, with a probability of $1-o\left(\frac{1}{T}\right)$, the inequality \newline $\left|\bar v_{t}-h_{i^*}(c')\right|\le\frac{bM}{3}$ holds for all $t\ge B$. Similarly,  $\left|\overline{|\tau_{t}|}-g_{i^*}(c')\right|\le\frac{LM}{3}$ also holds for all $t\ge B$. Since the provider can only inflate the length of the reported token sequence, we have $\overline{|\tau_{t}'|}-l'\le\frac{LM}3$.

Consequently, it can be inferred that with a probability of $1-o\left(\frac{1}{T}\right)$,

\begin{equation*}
\begin{aligned}
    \bar v_{t}-p_{i^*}\overline{|\tau_{t}'|}&\ge h_{i^*}(c')-p_{i^*}l'-\frac{(b+p_{i^*}L)M}3\\
    &=\bar u'-\frac{(b+p_{i^*}L)M}3.
\end{aligned}
\end{equation*}
This implies that the provider will not trigger the $\texttt{validated}$ flag (turning it to false) with a probability of $1-o\left(\frac{1}{T}\right)$. Therefore, the expected utility for provider $i^*$ in the exploitation phase is then given by $U_2\ge T_R(p_{i^*}l'-c'l')$, where ``2'' in the notation $U_2$ refers to the second phase.

Consider an ideal utility possibly achievable for provider $i^*$. The strategy is similar: For each query of the exploitation phase, it calculates a pair

\begin{equation*}
\begin{aligned}
    (c'',l'')=\argmax_{\substack{c\in[0,p_{i^*}], l\in[g_{i^*}(c),L]\\ h_{i^*}(c)-p_{i^*}l\ge \bar u'-\frac{(b+p_{i^*}L)M}3}}\{p_{i^*}l-cl\}.
\end{aligned}
\end{equation*}

Then it incurs a cost of $c''$ and generates a corresponding token sequence $\tau''$. If $|\tau''|\le l''$, it reports an output token sequence of length $l''$; otherwise, it reports the truthful token sequence. Should the provider be sufficiently fortunate, this strategy would also allow it to avoid triggering the $\texttt{validated}$ flag and thereby achieve an ideal utility. Although the realization of such a scenario is probabilistically negligible, the resulting ideal utility serves as a definitive upper bound for the true optimal utility. Consequently, this ideal utility can be leveraged to compute a more strict additive approximation ratio. Hence, we know the ideal provider utility $U_2'=T_R(p_{i^*}l''-c''l'')$, and the additive approximation ratio $
U_2'-U_2=T_R[p_{i^*}(l''-l')-(c''l''-c'l')]$.

By Lemma~\ref{lem:gap}, we know that $U_2'-U_2$ is $O(T_RM)=O(T^{1-\epsilon}\log T)$.

Therefore, the strategy for provider $i^*$ in the exploitation phase (with the consideration of the blind trust phase I) is an $O(T^{1-\epsilon}\log T)$-approximate dominant strategy.

Finally, we analyze the exploration phase. We will prove that incurring the truthful cost and output token sequence in this phase is a dominant strategy for provider $i$, regardless of the strategies of other providers.

Let provider $i$ incurs an average cost $\bar\alpha_i$ and misreports an average extra token sequence length of $\bar\beta_i$. We know that $\bar\alpha_i\in[0,p_i]$, and $\bar\beta_i\in[0,L]$. When $i\ne i^*$, its total expected provider utility is

\begin{equation*}
\begin{aligned}
U_{\text{all}}\left(\bar\alpha_i,\bar\beta_i\right) &= Bp_i\left(\bar\beta_i+g_i(\bar\alpha_i)\right)-B\bar\alpha_ig_i(\bar\alpha_i) + Bp_iL\\
&+Bp_iL\left(3+ \frac{2h_i(\bar\alpha_i)}{p_iL}-\frac{2\left(\bar\beta_i+g_i(\bar\alpha_i)\right)}L\right)^+.
\end{aligned}
\end{equation*}

We denote $G=3+ \frac{2h_i(\bar\alpha_i)}{p_iL}-\frac{2\left(\bar\beta_i+g_i(\bar\alpha_i)\right)}L$. If $G\ge0$, we first take the partial derivative with respect to $\bar\alpha_i$, and we have
\begin{equation*}
\begin{aligned}
\frac{\partial U_{\text{all}}}{\partial \bar\alpha_i}\left(\bar\alpha_i,\bar\beta_i\right) &=Bp_i\frac{\dif g_i}{\dif\bar\alpha_i}(\bar\alpha_i)-B\bar\alpha_i\frac{\dif g_i}{\dif\bar\alpha_i}(\bar\alpha_i)-Bg_i(\bar\alpha_i)\\
&+Bp_iL\cdot\left(\frac1{p_iL}\cdot\frac{\dif h_i}{\dif\bar\alpha_i}(\bar\alpha_i)\vphantom{-\frac{\frac{\dif g_i}{\dif\bar\alpha_i}(\bar\alpha_i)}L}\right.\\
&+\left.\frac1{p_iL}\left(\frac{\dif h_i}{\dif\bar\alpha_i}(\bar\alpha_i)-p_i\frac{\dif g_i}{\dif\bar\alpha_i}(\bar\alpha_i)\right)-\frac{\frac{\dif g_i}{\dif\bar\alpha_i}(\bar\alpha_i)}L\right)\\
&=B\left(\frac{\dif h_i}{\dif\bar\alpha_i}(\bar\alpha_i)-\bar\alpha_i\frac{\dif g_i}{\dif\bar\alpha_i}(\bar\alpha_i)-g_i(\bar\alpha_i)\right)\\
&+B\left(\frac{\dif h_i}{\dif\bar\alpha_i}(\bar\alpha_i)-p_i\frac{\dif g_i}{\dif\bar\alpha_i}(\bar\alpha_i)\right)\\
&\ge0.
\end{aligned}
\end{equation*}

On the other hand, the partial derivative with respect to $\bar\beta_i$

\begin{equation*}
\begin{aligned}
\frac{\partial U_{\text{all}}}{\partial \bar\beta_i}\left(\bar\alpha_i,\bar\beta_i\right)&=Bp_i-2Bp_i<0.
\end{aligned}
\end{equation*}

Similarly, when $\bar\beta_i=0$ (which means reporting truthful token sequences at all times), $U_{\text{all}}$ takes its maximum value ($\bar\alpha_i$ fixed).

Since $U_{\text{all}}$ is additively separable, when $\bar\alpha_i=p_i$ and $\bar\beta_i=0$, $U_{\text{all}}$ reaches its maximum value

\begin{equation*}
Bp_iL+Bp_iL\left(3+ \frac{2\mu_i^r}{p_iL}-\frac{2\mu_i^l}L\right).
\end{equation*}

We know that $3+ \frac{2\mu_i^r}{p_iL}-\frac{2\mu_i^l}L\ge1$, so $G\ge0$ holds.

If $G<0$, then
\begin{equation*}
\frac{\partial U_{\text{all}}}{\partial \bar\beta_i}\left(\bar\alpha_i,\bar\beta_i\right)=Bp_i\ge0.
\end{equation*}

Therefore, when $\bar\beta_i=L$, $U_{\text{all}}$ reaches its (ideal) maximum value of
\begin{equation*}
B(p_i-\bar\alpha_i)g_i(\bar\alpha_i) + Bp_iL\le2Bp_iL.
\end{equation*}

Since $3+ \frac{2\mu_i^r}{p_iL}-\frac{2\mu_i^l}L\ge1$,

\begin{equation*}Bp_iL+Bp_iL\left(3+ \frac{2\mu_i^r}{p_iL}-\frac{2\mu_i^l}L\right)\ge2Bp_iL.
\end{equation*}

Thus, the overall maximum value of $U_{\text{all}}$ is attained at $\bar\alpha_i=1$ and $\bar\beta_i=0$.

When $i=i^*$, since the utility of the exploitation phase is independent of both $\bar\alpha_i$ and $\bar\beta_i$, the maximum value is also achieved when $\bar\alpha_i=1$ and $\bar\beta_i=0$. Thus, in either case, incurring the truthful cost and reporting the truthful token sequence is an $O(T^{1-\epsilon}\log T)$-dominant strategy for all providers.

\end{proof}

\thmuti*

\begin{proof}
\begin{equation*}
\begin{aligned}
    \text{user utility} &=\mathbb E\left[\sum_{t=1}^T\left(v_{d_t,t}-p_{d_t} \cdot |\tau_{d_t,t}'|\right)\right]\\
    &\ge\sum_{i=1}^K(\mu_i^r-p_i\mu_i^l)B+(\bar u'-\frac{(b+p_{i^*}L)M}{3})T_R\\
    &+\sum_{i=1}^K(\frac{2\mu_i^r}{p_iL}-\frac{2\mu_i^l}L+3)p_iBL\\
    &=u_{SB}-O(T^{1-\epsilon}\log T+T^{2\epsilon}).
\end{aligned}
\end{equation*}
\end{proof}

\end{document}